# Retail Analytics in the New Normal: The Influence of Artificial Intelligence and the Covid-19 Pandemic

Yossiri Adulyasak
Associate Professor of Operations Management
HEC Montreal
3000, Chemin de la Cote-Sainte-Catherine, Montreal, H3T 2A7, QC, Canada.
yossiri.adulyasak@hec.ca

Maxime C. Cohen
Professor of Operations Management
Desautels Faculty of Management, McGill University
1001 Rue Sherbrooke O., Montreal, H3A 1G5, QC, Canada.
maxime.cohen@mcgill.ca

Warut Khern-am-nuai
Associate Professor of Information Systems
Desautels Faculty of Management, McGill University
1001 Rue Sherbrooke O., Montreal, H3A 1G5, QC, Canada.
warut.khern-am-nuai@mcgill.ca

Michael Krause
Director of Strategy
Aspen Technology
2000 Rue Mansfield Suite 1100, Montreal, H3A 1M4, QC, Canada
michaelmergykrause@gmail.com

## Abstract

The COVID-19 pandemic has severely disrupted the retail landscape and has accelerated the adoption of innovative technologies. A striking example relates to the proliferation of online grocery orders and the technology deployed to facilitate such logistics. In fact, for many retailers, this disruption was a wake-up call after which they started recognizing the power of data analytics and artificial intelligence (AI). In this article, we discuss the opportunities that AI can offer to retailers in the new normal retail landscape. Some of the techniques described have been applied at scale to adapt previously deployed AI models, whereas in other instances, fresh solutions needed to be developed to help retailers cope with recent disruptions, such as unexpected panic buying, retraining predictive models, and leveraging online-offline synergies.

## Retail Analytics in the New Normal: The Influence of Artificial Intelligence and the Covid-19 Pandemic

The COVID-19 pandemic has reshaped the retail landscape, compelling businesses to adopt innovative technologies. A prime example is the explosion of online grocery orders and the technologies employed to streamline this process. This disruption served as a stark awakening for many retailers, prompting them to recognize the transformative power of data analytics and artificial intelligence (AI). Simultaneously, the pandemic has also driven significant shifts in consumer behavior and societal norms. Collectively, these adaptations have given rise to the "new normal" of the post-COVID era. In this article, we explore the opportunities that AI presents to retailers in the new normal retail landscape, applying the fundamental principles of retailing science presented by Fisher et al. [1, 2] as the backdrop. Some of the techniques described have been applied at scale to adapt previously deployed AI models, whereas in other instances, fresh solutions needed to be developed to help retailers cope with recent disruptions, such as unexpected panic buying, retraining predictive models, and leveraging online-offline synergies.

**Essential Retail: Disruption and New Normal**

Since the emergence of the COVID-19, essential retailers such as supermarkets and pharmacies have experienced unforeseen changes to their business. Lockdown restrictions and curfews have upended the daily routine of hundreds of millions of individuals, resulting in fundamental shifts in consumer behavior. As people spend more time at home, demand for groceries (see Figure 1), alcohol, and personal-care products has sharply increased, whereas demand for attire, cosmetics, and sunscreen has weakened [3]. In addition to changing consumers' spending habits, the way that consumers purchase goods has also evolved. In North America, e-commerce sales have increased by more than 30% in 2020 [4]. This increase has been partially driven by the rapid emergence of online grocery orders. In addition, a prolonged economic recovery will likely further affect consumer behavior in the years to come [5].

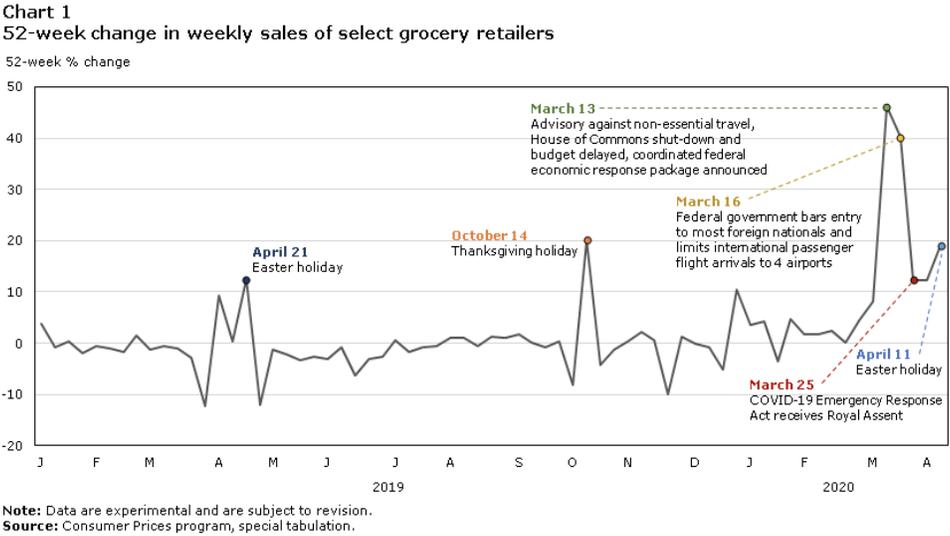

Figure 1: Weekly sales of select grocery retailers in 2019-2020 (source: Statistics Canada).

Besides the above "rationalizable" changes in consumer behavior, the year 2020 has seen several unprecedented widespread, unexpected panic buying waves. Hand sanitizers, masks, gloves, and disinfectants were the first group of products experiencing empty shelves, followed by toilet paper, baking supplies, and shelf stable foods [6]. As North America continues to battle COVID-19 and its newly emerging strains, there have been reports of widespread stockouts [7], highlighting that past panic buying waves were not insular events. In fact, drastic shifts in demand and consumer behavior are not the only challenges faced by retailers. The COVID-19 global pandemic has also caused significant surges in shipping costs, shortages of supply and labor, which make it exceedingly difficult for retailers to cope with such an unprecedented disruption [8, 9]. In this article, we focus on the challenges faced by retailers at the customer end, and not on supply chain and distribution challenges.

For essential retailers, keeping up with this new normal of rapidly shifting demand patterns poses great challenges. First, customers' livelihood depends on the accessibility of food, basic staples, and medicines making vulnerable and mobility-impaired individuals disproportionately affected by low availability in their local stores. Second, empty shelves may incentivize customers to mobilize more to complete their shopping list, which may further spread the virus. And last, stockouts and (perceived) shortages often lead to customer frustrations that may damage retailers' reputations [10]. In addition to those consequences, natural disasters and unexpected events (e.g., wildfires and widespread national protests) can exacerbate these trends, as more people are confined to their houses and feel a sense of incertitude, which is a key driver for panic buying [11].

Traditional retail planning processes, which were primarily designed to achieve cost efficiency, could not proactively react to sudden disruptive shifts caused by such black swan events [12]. As suggested in [1], data-driven retail analytics can be used to alleviate such issues and improve various aspects of retailers, including forecasting, inventory planning, and supply chain management. In this article, we argue that the combination of data, both internal and external, and machine learning (ML) and AI techniques[1] can further improve the operations and mitigate the effects of rapid demand shifts, which can eventually protect retailers' customers and reputation.

---

[1] Examples of AI-based implementations can be found in the literature [13-17].

For example, in terms of inventory planning, research has demonstrated that the adoption of AI-based systems to help plan inventory and fulfillment of essential products has helped increase the in-stock rate at JD.com by 50% in February 2020 following the first pandemic outbreak [13]. In addition, more than 100,000 of their industrial retail partners have also achieved a 100% improvement in turnover days year-over-year following the use of AI-based systems developed by JD.com. Relatedly, in a recent study [14], a data-driven prescriptive analytics tool could help the retailer increase access to essential products by approximately 57% amid a strong panic buying behavior.

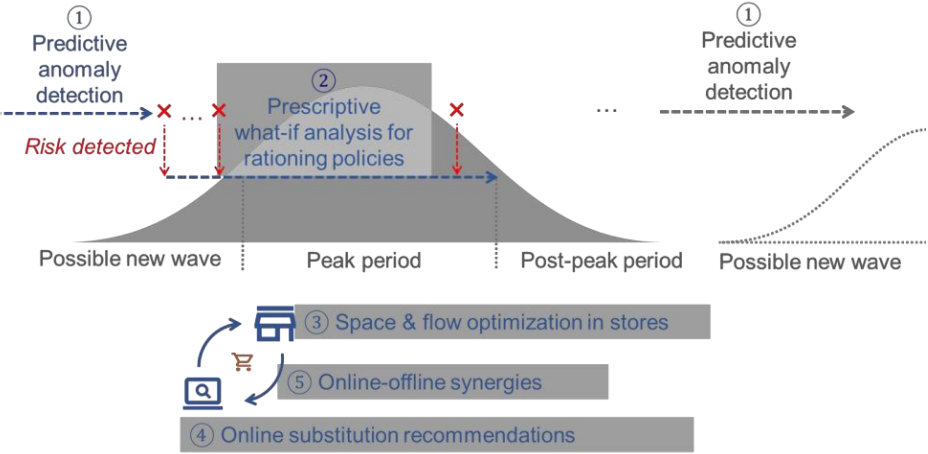

Figure 2: Illustration of data and AI initiatives for essential retailers at different phases of the pandemic.

The illustration of the main data and AI initiatives discussed in this article is shown in Figure 2. We start by discussing how real-time demand anomaly detection can be instrumental to detect demand disruptions during the initial stage prior to the arrival of a full panic buying wave. We then explain how a counterfactual what-if scenario tool can be used once risks (i.e., pertinent anomalies) are detected to simulate different panic buying mitigation approaches, and how external data can be strategically leveraged for retail planning. Next, we discuss how retailers need to rethink their approaches to keep existing

AI models up to date in a continuously disrupted landscape as we have all witnessed since the first pandemic wave. We detail how model retraining needs to become more dynamic and routinely performed. In this context, we outline how retail data scientists can define training datasets that sufficiently reflect current consumer trends to yield highly performing AI models. Finally, we discuss how essential retailers can adjust their omni channel operations to effectively serve customers in both brick-and-mortar stores and online. More specifically, we describe how changes to physical stores' layouts due to social distancing guidelines can be leveraged using AI recommendations to increase basket sizes, how online order fulfillments can be optimized, and how retailers can exploit synergies between their physical and online channels.

**Leveraging Data and AI to Detect Shifting Retail Trends**

A key challenge faced by essential retailers in early 2020 was that they could neither anticipate nor prepare for the drastic unforeseen changes in consumer demand, making forecasting based on historical data and data-driven supply chain management particularly challenging [18]. In addition to a sudden unexpected increase in demand for groceries [19], waves of panic buying in March and November 2020 were especially challenging for retailers. To proactively deal with subsequent panic buying waves, we argue that it is desirable for retailers to develop an AI-based approach that can successfully detect shifts in consumer demand as early as possible to allow for a timely deployment of mitigation measures, which eventually helps with the overall demand forecasting framework. The challenge in developing such a solution is to make it flexible enough to be readily adaptable to detect different types of trends.

*Anomaly Detection*

Oftentimes, retailers realize the presence of demand anomalies when it is too late to react and alleviate the impact of the disruption. Rather than relying on simple outlier rules, retailers can use AI models that monitor transaction-level sales data in real-time to detect subtle demand anomalies that appear during the initial onset of panic buying. For example, an unsupervised deep learning-based approach, notably the variational autoencoder, can be used for an early detection of panic buying [14]. Unlike traditional approaches in anomaly detection, which are mainly applicable to univariate time-series data, this type of AI-based approaches can be used to deal with multivariate time-series data collected from various sources without requiring extensive feature engineering processes. Once detected, essential retailers have two key levers to mitigate its effects: (i) replenishing products at a higher rate (when possible), and (ii) restricting the number of items that a single customer can buy (called a rationing policy). Both levers are most effective when applied early, when there is still sufficient stock in stores and distribution centers. If such anomalies can be detected reliably and promptly, retailers can "buy themselves some time" by restricting the number of items per customer to stretch available supply until new sources of supply are found.

Since monitoring a large stream of real-time data can be costly, retailers may select a subset of products or stores to serve as early-warning signals for panic buying. These subsets are typically chosen by retailers based on past experience. The goal is to focus on products and stores that can be seen as leading indicators, which can be effectively selected by analyzing the patterns leading to disruptions from historical purchase data [14, 20]. In addition, once an AI model is developed and calibrated, one can apply it to a multitude of product categories and stores to assess (and remove) the presence of potential biases. For example, urban stores are likely to experience panic buying prior to rural stores. Once anomalies in the subset of products or stores are detected, the AI model, which is trained using several time-series data and product attributes, can effectively be used to monitor and detect anomalies potentially leading to product availability issues.

Anomaly detection is not limited to detecting panic buying. Specifically, alternative, more subtle, shifts in consumer behavior can also be detected. As consumers spend more time at home, new eating trends have emerged. For example, a large number of consumers started baking and pickling in 2020 [21], so that retailers are likely to observe demand spikes for specific types of products. Demand anomaly detection can generally be used to detect such shifts in consumer behavior and help retailers adjust inventory decisions and promotion strategies accordingly.

*What-if Scenario Tool*

To ensure that retailers can act upon the detected anomalies, they may use a counterfactual mitigation approach simulation tool to help them identify products with a high risk of stock-out and evaluate a menu of mitigation approaches. Such a tool allows retailers to simulate at the store (or region) level how long the current stock of products targeted by panic buying (or by other demand anomalies) would last. This information, combined with the replenishment calendar, will allow the retailer to identify which products have a high stock-out risk and require immediate attention, which ultimately improves the data-driven supply chain management system. As a second functionality, the simulation tool can help evaluate different mitigation approaches, such as identifying the right rationing policy (i.e., imposing a limited number of items per customer). This allows retailers to strike a better balance between allowing customers to purchase their desired quantities and ensuring fair distribution.

*Leveraging External Data*

As consumers spend an increasing amount of time online and, specifically, on social media, retailers can leverage external data feeds to detect rapidly emerging trends and adjust their strategic planning. This can be done at the category level or at the specific brand/product level. For example, grocery retailers can monitor social media activities and search trends for keywords related to recipes and hashtags, as a majority of consumers research recipes online [22]. Retailers can then increase their demand forecast for products related to the trending recipes, adapt their promotion strategies, and explore augmenting their assortment in the relevant categories. As social media engagement on grocery-related posts has significantly increased amid the COVID-19 pandemic, retailers can gauge which brands and products are likely to become trendy following a specific post and can strategically adjust their planning [23].

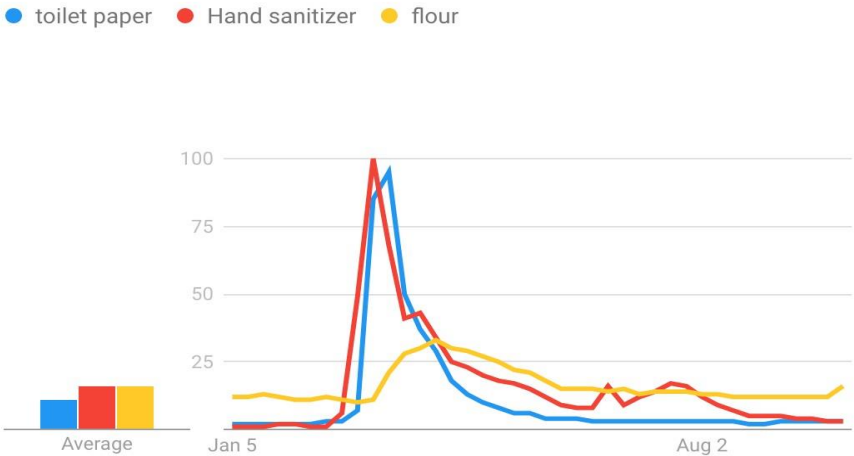

Figure 3: Google Trends for products targeted by panic buying.

In addition to the conventional use of social media data in a descriptive fashion, retailers can use these external data to augment their AI models. In the context of anomaly detection, for example, retailers can leverage external data sources to increase the predictive performance in detecting pertinent anomalies. Interestingly, during the first North American panic-buying wave in March 2020, a surge of online searches and social media outrage for stocked-out products occurred. Such data is available via Google Trend (Figure 3 shows the spike of searches related to toilet paper, hand sanitizer, and flour during the March 2020 panic-buying wave).

**Proactive Measures to Keep AI Models Up to Date**

Data-driven retail analytics and the use of AI tools critically depend on accurate and available historical data [1]. However, as consumer behavior has fundamentally shifted since the beginning of the pandemic, the way that AI is used in retail needs to also be revisited [24]. AI models learn patterns from historical data and generate predictions and recommendations by applying these learned associations to future data. Thus, AI models, especially the ones trained on a large amount of historical data, can perform very well when the consumer behavior remains stable over time. In the current constantly evolving environment, however, this assumption is no longer realistic in many cases.

There are myriad examples that illustrate the need to reevaluate and retrain AI models. For instance, before the pandemic, online grocery deliveries were mostly used by technologically savvy urbanites. Any model trained on historical online sales data will base its predictions on the preferences of this specific customer segment. However, since the start of the pandemic, a broader proportion of the population has started using online grocery deliveries. As a result, a demand forecasting model trained on pre-pandemic data would not be able to accurately forecast the increase in demand, which is partially driven by a demographic shift.

Similar examples can be found in AI models based on in-store sales data. Compared to pre-pandemic times, people visit supermarkets less frequently but tend to buy larger baskets per visit. A promotion optimization model fed with basket-size data to prescript promotions could thus mistakenly assume that the increase in basket size is correlated to specific promotions and recommend repeating the same promotion agenda in the future.

As consumer priorities continue to shift, econometric models also require major adjustments. For example, price elasticity calculations used in the context of dynamic pricing aim at maximizing revenue or profit, while carefully balancing demand with supply. At the beginning of the pandemic, the demand for many products had instantly become inelastic. For instance, in the first pandemic wave, consumers had been purchasing hand sanitizers at unprecedented prices [25]. In this case, a pricing model trained on pre-pandemic data would not be able to capture the full revenue upside and would not accurately balance demand with supply. Decisions related to the ethics of spiking prices during challenging times is beyond this article's scope and has been the focus of heated media debates [26]. Similarly, as we are facing a slow economic recovery, consumers are likely to become more price sensitive post-pandemic. The previous examples highlight the intricacy of keeping models up to date. It is not sufficient to just retrain models with post-pandemic data and continue using them in the same way as before. Certain trends, such as the broad adoption of online grocery shopping, are likely to stay, whereas paying exorbitant prices for hand sanitizers is likely a temporary phenomenon.

Retailers that currently use AI methods need to monitor the models' performance in real time in order to know when to trigger a model retraining. The retraining process is known to be critical especially for AI-based systems deployed in a constantly varying environment where collected data continue to evolve [27]. Various recent techniques such as DeltaGrad [28] have been developed to specifically address this issue.

For models trained on sales data, retailers can repurpose an anomaly detection model as an indicator of when to retrain their AI models [14]. In addition to the use of such a model as a real-time detection mechanism, persistent or recurring anomalies for specific products, stores, or channels indicate that current demand patterns have deviated significantly from patterns present in the training data. Once recurring anomalies are detected, the relevant KPIs need to be compared to their values in the pre-anomaly time frame. For a dynamic pricing model, one could compare the sales velocity and inventory levels pre- and post-anomaly to assess the model's effectiveness in adjusting prices to evolving demand patterns. Once a significant drop in the model performance is detected, a retraining phase should be triggered to recalibrate the model parameters and learn the new reality. For example, such countermeasures have been effectively deployed at JD.com to improve the resilience of their supply chain planning process [13]. A continuous monitoring process through measures like thresholds based on forecasting accuracy and tracking signals can also be put in place to alert retailers when sudden shifts have been detected.

Another common issue, which is amplified as a result of sudden demand surges and limited supply, is censored demand data given that retailers have faced unprecedented levels of stockouts. This issue presents a significant challenge in the training process of demand forecasting models due to the fact that the recorded sales data will be very different from the (unknown) actual demand. To deal with this issue, retailers can make use of recently observed rates of sales in conjunction with choice models such as the multinomial logit model to estimate or simulate the unconstrained demand [29]. Apart from deciding the optimal time and frequency of model retraining, finding the time periods of the training dataset poses its own challenge. In the light of the pandemic disruption on consumer behavior, one might be tempted to exclude all pre-pandemic data and train models exclusively using post-pandemic data. This solution, however, might not be feasible, as many AI models require a large amount of training data. Furthermore, although consumer behavior has recently shifted substantially, a large portion of fundamental behavior remains the same. For example, consumers who buy hot-dogs are likely to concurrently buy buns, and baskets with pasta are more likely to include pasta sauce. Thus, it is important to strike a good balance between pre- and post-pandemic data to ultimately build a satisfactory training dataset.

As a first step, data scientists should divide historical data into different time frames, namely, a pre-pandemic time frame ending with the first occurrence of pandemic-induced panic buying, panic-buying time frames, and post-pandemic "steady state." In a second step, they need to determine the proportion of each time frame in the training data. This is an experimental and iterative process, where data scientists extract at random (i.e., subsample) a different number of observations from past time frames. In a third step, the randomly chosen observations are combined with post-pandemic steady-state data to create a balanced training dataset [30]. Then, one can measure the model's KPIs and readjust the training dataset until the models reach the desired performance level. A human-machine hybrid system that strategically leverages human interventions in estimating and adjusting important inputs of demand forecasts, such as product shares, customer arrival rates, and growth rates, can be a promising alternative to enhance the performance and reliability of AI models [31]. A summary of common issues related to training AI models using historical data amid the pandemic is presented in Table 1.

Table 1: Common issues related to training AI models using historical data amid the pandemic.

| Data issues amid the pandemic | Potential countermeasures and solutions |
|---|---|
| Data shifts due to changes in purchase behavior | • Uses of change point and anomaly detection algorithms<br><br>• Continuous monitoring process (e.g., through tracking signal measures) |
| Censored data due to limited inventory and demand surges | • Data pre-processing and simulation using sale-rate estimation<br><br>• Unconstrained demand adjustments |
| Insufficient training data | • Data augmentation and implementation of ML models trained across multiple items and stores (e.g., transfer learning)<br><br>• Frequent model retraining<br><br>• Use of human-machine hybrid models |

**Winning in an Omni-Channel World in the New Normal**

Stay-at-home orders in the Spring of 2020 have started an unprecedented trend in demand for online grocery orders (see Figure 4). This represents a challenge to many essential retailers who are brick-and-mortar native and typically focus on operations and supply chain management related to in-store customer experience. Retailers can leverage AI to capitalize on the increased demand of online grocery orders. They can also update their physical stores' layouts to better cope with the fundamental changes in shopping behavior amid the pandemic. In this section, we show how adjusting physical stores' layouts to conform with physical distancing rules can be an opportunity to optimize planograms. We also discuss how prescriptive AI models can be used to integrate online order fulfillment more effectively to day-to-day operations, and how retailers can leverage online-offline synergies to adjust their strategy in the post-pandemic new reality.

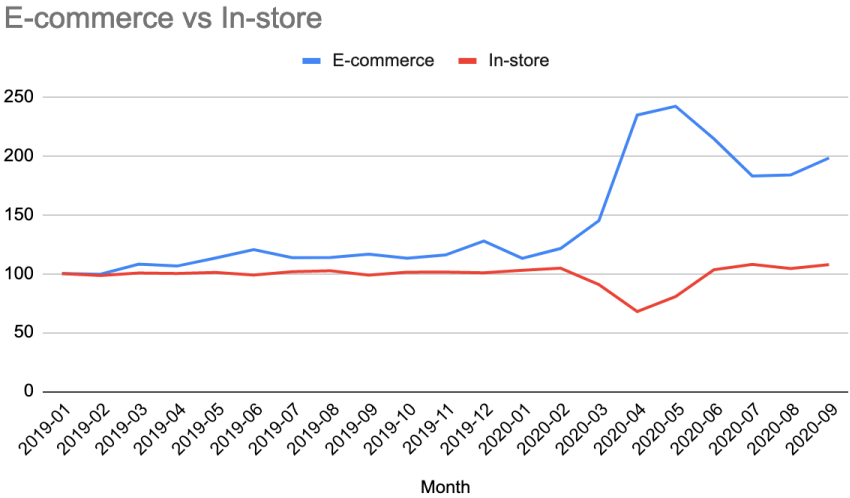

Figure 4: Indexed monthly e-commerce versus in-store grocery sales (source: Statistics Canada; Jan. 2019 = 100).

*Adjusting Physical Stores*

Although the adoption of online grocery shopping has skyrocketed during 2020, the physical supermarket is here to stay. First, the physical format still offers value in capturing demand that cannot be effectively met by the online channel (e.g., last-minute purchases). The value of physical stores has even been recognized by Amazon, which, in addition to its Whole Foods locations, is building three grocery and convenience store banners [32]. Second, even if grocers wanted to give up their physical stores, many of them cannot do so, as they are bound by long-term lease agreements. So, in addition to optimizing their online operations, it is vital for retailers to adjust their physical stores to the post-pandemic landscape.

In the first COVID-19 wave, supermarkets have been identified as the most common virus exposure location in certain jurisdictions [33]. To minimize in-store virus spread, many retailers have redesigned certain aspects of their stores. As the store floor plan is set in many cases and since any additional space for consumer traffic comes at a loss of merchandising space, several retailers such as Walmart have implemented one-way directional paths in their stores [34]. Having a unidirectional customer flow along the aisles brings a new opportunity for retailers to re-optimize their store layout, shelving configurations, and planogram decisions. As customers are supposed to follow predetermined paths, the sequence by which they pass products is also determined. Retailers can thus leverage data-driven methods to develop a planogram optimizer that accounts for demand transference (i.e., which alternative choices consumers make when their first desired option is out of stock), cross-sales (i.e., which products are bought to complement one another), and halo effects (i.e., which products benefit from positive consumer perceptions) [35].

*Order to Fulfillment*

A common issue in online grocery shopping occurs when products that have been ordered online are out of stock in the store. In that case, the picker either needs to contact customers and ask which alternatives they would like or make the choice themselves and confirm with customers later. In many cases, selected substitutions will not satisfy customers' needs [36]. Requiring pickers to make judgment calls on substitutions in real time or to check with customers is disruptive to the picking process and can cause significant losses in terms of refunds and customer satisfaction.

To better satisfy customers' needs and streamline the picking process, AI-based methods can be used to develop a substitution recommendation engine. Demand transference models aim to quantify substitution behavior in cases where the primary desired product is out of stock. Namely, it answers the following question: If a specific product is out of stock, what is the additional generated demand for similar products in the same category? Since substitution behavior cannot be directly observed in the sales data, ML-based approaches, particularly Markov Chain Monte Carlo (MCMC) [37] or even an ensemble method consisting of various ML models [38], can be leveraged to estimate the latent variables that capture demand transference. These models can either be trained using in-store transactions data or online data (searches, clicks, and purchases). The estimated demand transference weight between a pair of products represents the likelihood of substitution, which can be computed for each store and each customer segment. If the training data can be enhanced with customer loyalty information, substitution patterns can be determined at a more granular level (e.g., groups of customers that follow a similar behavior), which can further improve the quality of substitution recommendations. The implementation of such an ML-based system at Alibaba has enabled a 28% increase in revenue per customer [38]. In practice, online customers can also be presented with customized alternative options for out-of-stock products or receive an automated text message to accept/reject a specific substitution when the picker encounters a stockout.

*Leveraging Online-Offline Synergies*

Following a substantial growth in online orders, many retailers started using their stores as additional local fulfillment centers [39]. Consequently, flow optimization in stores has become more difficult than ever before due to the significant presence of store pickers during operating hours. Motivations behind this practice include removing the need for building a parallel online supply chain, leveraging client proximity when supermarkets are located in urban areas, and leveraging additional organizational benefits. AI-methods can help retailers maximize pickers' efficiency, while minimizing the impact on store operations. To alleviate undesirable customers' disruptions, retailers can further leverage real-time traffic, basket and online order data to dynamically schedule the in-store picking tasks. To extract such information for the flow optimization process, shoppers' purchase intentions and paths for each individual store can be learned through a tree-based learning model in conjunction with clustering techniques [40], which allows the retailer to understand the purchase patterns of different customer segments during different times of the day and schedule in-store pickers to minimize undesirable customers' disruptions.

In the same vein, retailers can exploit advanced routing mechanisms to identify the optimal picker's trajectory and the picking sequence through the store. These instructions can then be displayed on the picker's tablet. These routes can be optimized for picking efficiency, while respecting routing constraints of one-way aisles as well as accounting for other pickers' routes. Approaches to optimize picking orders have been extensively studied in the context of warehouses [41].

Many retailers also use micro-fulfillment centers (MFCs) to fulfill online orders so as to reduce inefficiencies and extra work in-store. Nevertheless, the majority of online orders cannot be fully fulfilled from the MFCs due to space constraints and to the limited assortment of products, leading to unnecessary trips to the stores to complete such orders. To improve the order coverage of the MFCs, retailers can apply association rules mining algorithms [42] to online purchase data and decide which products and how many units should be stored at the MFCs. In addition, during an online shopping session, retailers can further leverage the predictive sequential association rules [43], as an online recommendation algorithm to sequentially recommend products that are available at the MFC, which likely satisfy customer's needs and allow the entire order to be completed from a single MFC location.

An alternative approach to leverage online-offline synergies is by incentivizing customers to visit a physical store and pick up their order at times when deliveries are reaching capacity. Retailers can leverage machine learning models to accurately predict the required delivery time for each online order. Several retailers leverage ML-based prediction by using a combination of supervised learning techniques with different levels of model complexity ranging from regressions to neural networks to estimate the total customer delivery time using order information (e.g., size, time of delivery), routing information (distance from MFC to customer), and external factors (e.g., weather, traffic). Having a better view on delivery time is estimated to increase the total possible number of deliveries in a given year by 10-15%, whereas the optimal vehicle routing yields an efficiency increase of last mile delivery of 10% [44].

If the predicted delivery time surpasses an acceptable level, consumers can be incentivized to visit a specific store to pick up their order. Grocery retailers have collected vast amounts of consumer data through loyalty programs and have leveraged this data to design personalized offers and promotions. By knowing which type of incentives (e.g., coupons for certain products) are most likely to influence a specific behavior, supermarkets can decrease the cost of incentivization while effectively influencing its customers to visit the store to pick-up their delivery. A wide range of consumer nudging have been explored to incentivize other habits, such as healthy eating [45].

A recent example of an online-offline synergy is Walmart's initiative to upgrade 200 stores to a new format that aims at encouraging in-store app usage [46]. Specifically, consumers can use the smartphone application to navigate the store, scan products to check availability and compare similar products, and pay via a self-check-out process. Several other retailers are following suit and started offering different alternatives for a hybrid customer experience, where the offline and online worlds are intertwined.

The COVID-19 pandemic has had a significant and continued impact on retailers, forcing the acceleration of advanced analytics and AI initiatives due to changing consumer behavior, as well as highly uncertain business and operational environments. Such transformations are inevitable to increase the competitiveness of retailers in the new normal, which is still constantly evolving. Retailers will need to rely more heavily on data-driven capabilities and proactively react to sudden shifts caused by unknown and disruptive uncertainties, while keeping in mind that the underlying models they rely on also require monitoring, updating, and retraining. With recent advances in business analytics and AI as well as process standardization capabilities such as MLOps in cloud platforms [47], such data-driven transformations can be achieved in a very cost effective manner [48]. Specifically, adopting and scaling AI models, like the ones discussed in this article, is now relatively straightforward and can be done with a low amount of resources and a small team of scientists and engineers. Essential retailers that capitalize on this AI revolution will have a significant advantage in the post-pandemic world.

## Summary: Materials & Methods

This article describes several AI techniques that can help leverage retail analytics in the new normal post-pandemic world. This section concisely summarizes these techniques.

Table 2: Summary of material and methodologies of AI models discussed in this article.

|  | Research type | Study area | Data source | Sample size |
|---|---|---|---|---|
| [13] | Observational study | Supply chain planning and inventory fulfillment during the pandemic | JD.com | Daily transaction data from January to April 2020 |
| [14] | Observational study | Demand anomaly detection during the pandemic | Grocery store chain in North America | Point-of-sale data from 42 metropolitan stores from January 1, 2018 to May 1, 2020 |
| [28] | Model development and computer simulation | AI model that can retrain with post-pandemic data | 4 publicly available datasets | More than 11,000,000 samples |
| [31] | Model development | Human-AI synergy in model evaluation | Partner company | More than 6,000,000 entries |
| [38] | Randomized field experiment | Model to determine products to display on an online marketplace | Alibaba | 10,421,649 visits during a one-week-long period |
| [40] | Model development | Customer segmentation modeling | Retail stores in Europe | point-of-sale data from January 2012 to May 2013 |
| [41] | Survey | Warehousing optimization | - | - |
| [43] | Model development and computer simulation | Product recommendation based on association rules | 7 publicly available datasets | More than 70,000 product baskets |

## Conclusions

In recent years, Artificial intelligence (AI) has been widely used in several functions of our society, including the financial market [49], legal systems [50], and education [51]. In this article, we discuss how AI can help retail companies to adapt to the post-COVID era, including the role of AI in enhancing inventory and supply chain management and how AI can help enable omni-channel retailing. In that regard, our paper complements prior works on the intersection between AI and the COVID-19 pandemic in engineering management [52-55]. Nevertheless, it is important to note that while AI models undoubtedly benefit retailers in improving their data analytics procedures and enhancing their operational decisions, the current AI tools are still subjected to several limitations, which provide excellent avenues for future research. For instance, data analytics tools often fail to provide accurate information when there is a lack of precise data, or when unforeseen events like pandemics or geopolitical tensions unfold. The rise of explainable and interpretable AI tools that allow users to observe the prediction processes and the details/interpretations of the models would be an impactful improvement in that regard [56]. In addition, most AI tools still require certain levels of human involvement, so that future research could aim to identify the optimal human-AI augmentation and how interactions between humans and AI tools could be leveraged to further enhance operational efficiency.